\documentclass[aps,prb,amsmath,amssymb,preprint,groupedaddress,showpacs]{revtex4}

\usepackage{graphicx}
\begin{document}


\title{Quantized Landau level spectrum and its density dependence.}


\author{Adina Luican}
\affiliation{Department of Physics and Astronomy, Rutgers University, Piscataway, NJ 08855 }
\author{Guohong Li}
\affiliation{Department of Physics and Astronomy, Rutgers University, Piscataway, NJ 08855 }
\author{Eva Y. Andrei}
\affiliation{Department of Physics and Astronomy, Rutgers University, Piscataway, NJ 08855 }

\begin{abstract}
Scanning tunneling microscopy and spectroscopy in magnetic field was used to study Landau quantization in graphene and its dependence on charge carrier density. Measurements were carried out on exfoliated graphene samples deposited on a chlorinated SiO$_2$ thermal oxide which allowed observing the Landau level sequences characteristic of single layer graphene while tuning the density through the Si backgate. Upon changing the carrier density we find abrupt jumps in the Fermi level after each Landau level is filled. Moreover, the Landau level spacing shows a marked increase at low doping levels, consistent with an interaction-induced renormalization of the Dirac cone.
\end{abstract}


\maketitle

One of the hallmarks of the relativistic charge carriers \cite{RevModPhys.81.109,abergel2010properties} in graphene is the appearance in a magnetic field of an unusual Landau level (LL) at zero energy which reflects the chiral symmetry of the low lying excitations. The presence of this LL has been inferred in magneto-transport measurements employing the standard configuration of graphene supported on SiO$_2$ \cite{novoselov2005two,zhang2005experimental} from the conspicuous absence of a quantum Hall plateau at zero filling-factor. Remarkably because in graphene the carriers reside right at the surface, the LLs (including the LL at zero-energy) can be accessed directly through scanning tunneling spectroscopy (STS) as was demonstrated in studies of graphene samples supported on  the surface of graphite \cite{PhysRevLett.102.176804,li2007observation}.  However, the LLs were not observed in STS measurements on graphene samples supported on insulating substrates which allow control of the carrier density through gating.  This is because due to the purely two dimensional nature of graphene,  substrate induced potential fluctuations obscure the intrinsic physics of the charge carriers close to the Dirac point. One way to overcome this limitation is to use suspended samples \cite{du2008approaching,bolotin2008ultrahigh} where transport measurements have shown that in the absence of the substrate the intrinsic Dirac point physics including interaction effects is revealed \cite{du2009fractional, bolotin2009observation}. The use of suspended samples is however limited due to their fragility, small size and reduced range of gating.  Finding a minimally invasive insulating substrate on which graphene can be gated and also visualized is therefore of great interest. 

By using scanning tunneling microscopy (STM) and spectroscopy (STS) we show that for graphene supported on SiO$_2$ substrates which were treated by chlorination to minimize trapped charges and in sufficiently large magnetic fields, the LL sequence specific to single layer graphene and its dependence on carrier density can be accessed. Upon varying the carrier-density sudden jumps of the Fermi-energy are observed after filling each LL. Moreover the measured density-dependence of the LL spacing shows a rapid increase upon approaching the Dirac-point, consistent with an interaction-induced renormalization of the Dirac cone.
\begin{figure}
\includegraphics[width=\columnwidth]{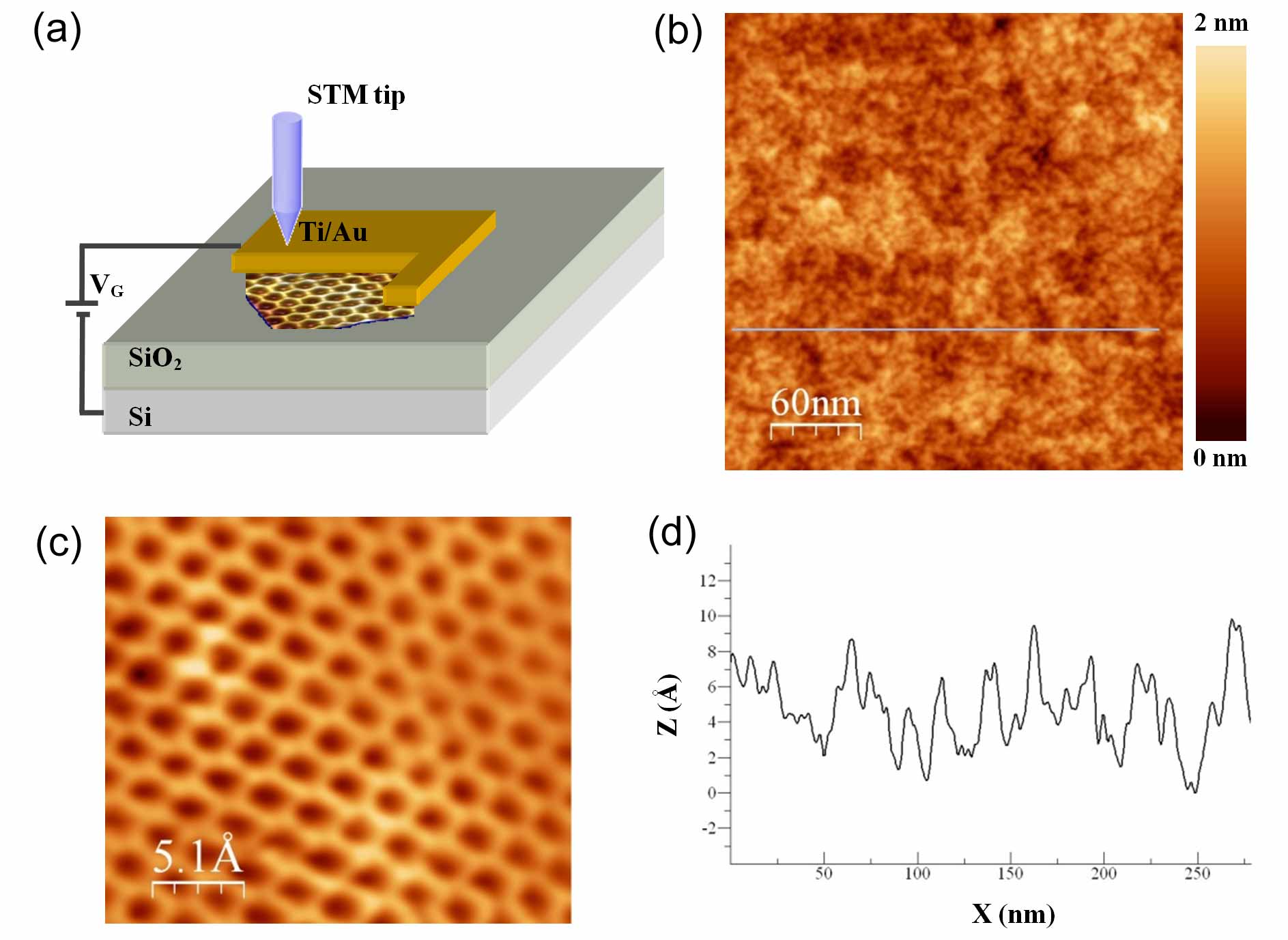}
\caption{\label{fig:fig1}(a) Sketch of the experimental set up showing a graphene flake deposited on Si/SiO$_2$, the STM tip above and the back gate connected to the graphene.(b) STM topography image of a typical 300 nm x 300 nm graphene area. Tunneling conditions: $I_t=20 \textrm{ pA}, V_{bias}=190 \textrm{ mV}$.(c) STM atomic resolution image on the graphene flake ($I_t=20 \textrm{ pA}, V_{bias}=300 \textrm{ mV}$).(d) Profile line through the topography image (at the position indicated in (b)) showing the corrugation of graphene on SiO$_2$ substrate.}
\end{figure}

A simple method for preparing high quality graphene samples is mechanical exfoliation from graphite followed by deposition on the surface of SiO$_2$ capping a Si crystal \cite{novoselov2005two}. This relative ease of sample preparation has stimulated numerous studies aimed at probing the extraordinary properties of graphene which range from novel physical properties to a wide range of applications. A similar method was used for the graphene devices studied here. Graphene samples prepared by exfoliation from HOPG graphite were deposited on the surface of a p-doped Si wafer capped with 285 nm of SiO$_2$. To minimize substrate induced disorder we used a chlorinated thermal oxide \cite{balk1988si} purchased from Nova Electronics which was annealed in forming gas at 230 $^{\circ}$C prior to the deposition of graphene. The metallic leads were patterned using e-beam lithography followed by evaporation of 1.5/30-40 nm Ti/Au. Prior to measurement, the samples were baked overnight in forming gas at 230 $^{\circ}$C. STM and STS measurements were performed at 4.4 K in a home built STM (Fig. \ref{fig:fig1}(a)). After baking, the surface of graphene is found to be free of fabrication residues on areas as large as 1 $\mu$m $\times$ 1 $\mu$m as measured by low temperature STM. A typical 300 nm $\times$ 300 nm area is shown in Fig. \ref{fig:fig1}(b). As found previously the samples are rippled on a fine scale of 2-5 {\AA} in the vertical direction and a few nm laterally \cite{PhysRevLett.102.076102,stolyarova2007high,ishigami2007atomic,deshpande2009spatially,zhang2009origin}. When zooming into atomic resolution (Fig. \ref{fig:fig1}(c)) the typical honeycomb structure observed is a first indication of the sample cleanliness and quality.

\begin{figure}
\includegraphics[width=\columnwidth]{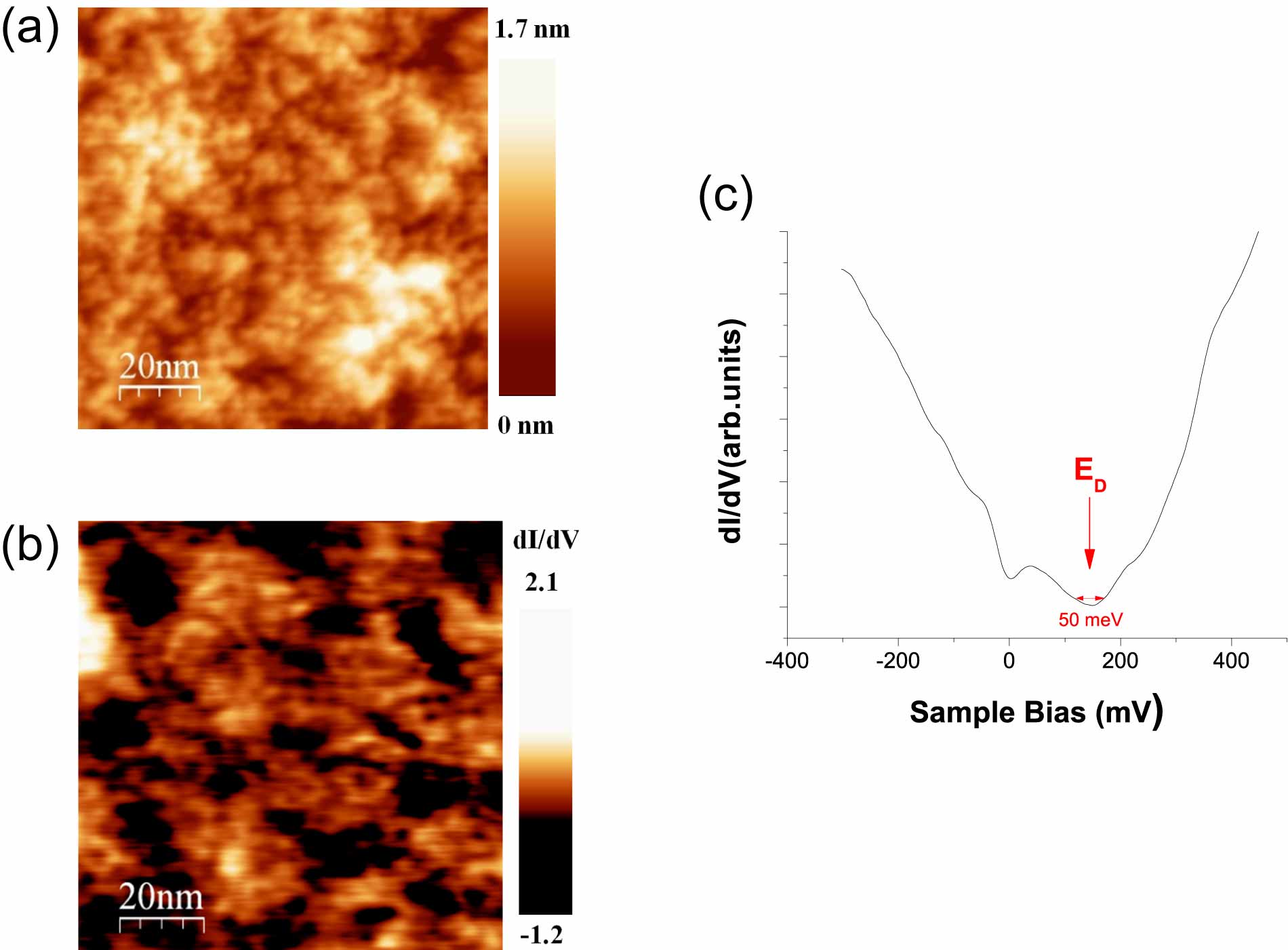}
\caption{\label{fig:fig2}(a) Topography image corresponding to the Local Density Of States (LDOS) map in (b) $(I_t=20 \textrm{ pA}, V_{bias}=140 \textrm{ mV})$. (b) LDOS map taken at E=140 meV showing doping inhomogeneity in graphene. (c) STS spectrum at B=0T showing smearing of the Dirac Point (indicated by ED) due to disorder.}
\end{figure}

The intrinsic density of states (DOS) of graphene in the absence of field is known to be V-shaped and to vanish at the Dirac point. In samples where substrate induced disorder is negligibly small: graphene flakes on graphite, it was possible to measure this intrinsic spectrum in zero field \cite{PhysRevLett.102.176804}. However, when graphene is deposited on an insulating substrate such as SiO$_2$ its properties are modified due to defects, strain, inhomogeneous doping and trapped charges \cite{zhang2009origin,teague2009evidence,martin2007observation}. These altering effects are responsible for a wide range of non-intrinsic spectra reported in the literature for graphene samples deposited on SiO$_2$ substrates \cite{zhang2009origin,teague2009evidence,geringer2010electrical}. In our case, scanning tunneling spectra in the absence of field and for a neutral gate show two dips, one  at the Fermi energy (E$_F$) where E=0 eV, and the other at $\approx$ 170 meV (Fig. \ref{fig:fig2}(c)). The deviation from the expected V-shape is attributed to substrate-induced disorder which smears the Dirac point. Similar to results reported in \cite{deshpande2009spatially,geringer2010electrical} but unlike those in \cite{zhang2009origin}, we do not see evidence for a the opening of a gap at E$_F$.

\begin{figure*}
\includegraphics[width=\textwidth]{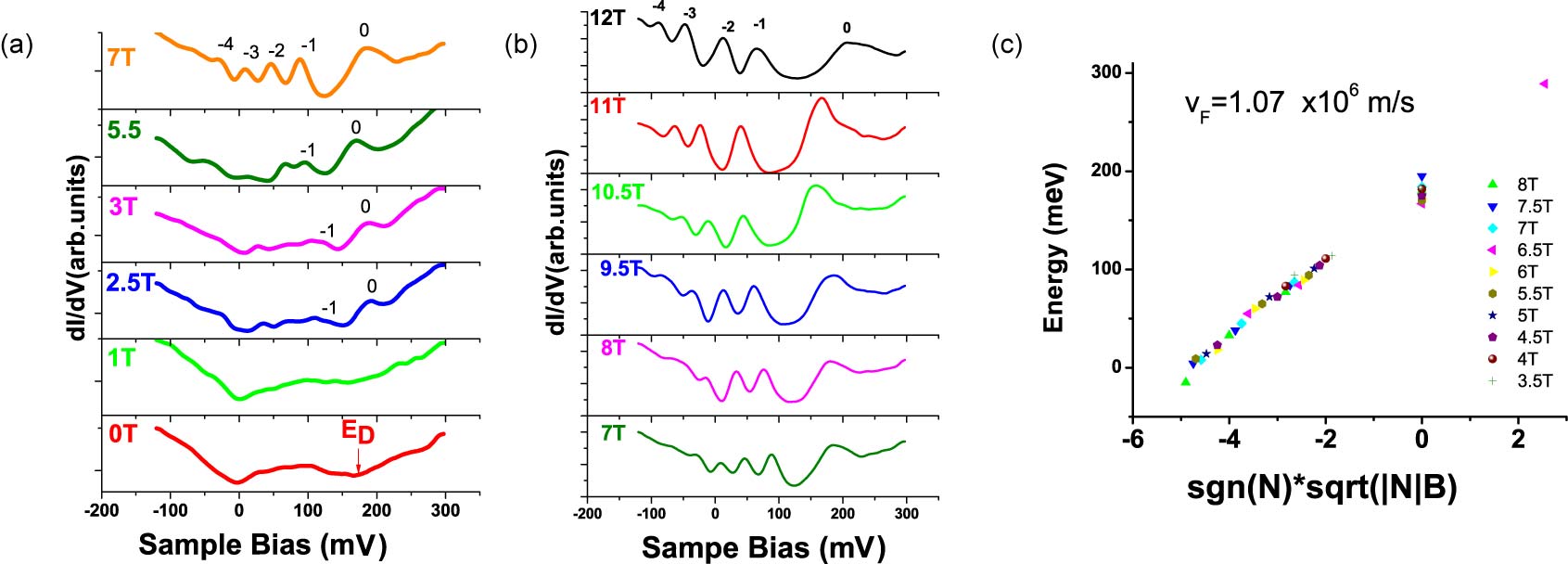}
\caption{\label{fig:fig3}Landau levels in ungated graphene on SiO$_2$ (V$_G$=0 V). (a) Scanning Tunneling Spectroscopy data in low fields B=1 T-5.5 T together with B=0 T and B=7 T data for comparison. (b) Landau Levels in higher fields B=7 T to B=12 T clearly resolving LL with indices N=-4, -3, -2, -1, 0. (c) Energy of the Landau Levels as function of $sgn(N)\sqrt{\left|N\right|B}$. From the linear dependence, specific of Dirac fermions the Fermi velocity is extracted. }
\end{figure*}

The inhomogeneous carrier density seen as electron-hole puddles close to the Dirac Point (DP) \cite{zhang2009origin,martin2007observation} is clearly visible  in the local density of state map shown in Fig. \ref{fig:fig2}(b) that corresponds to the region whose topography is shown in Fig. \ref{fig:fig2}(a).  We estimate a typical puddle size of d=20 nm. This imposes a length scale and a corresponding energy scale which separates between disorder-controlled and intrinsic phenomena.  In finite field the intrinsic physics of the charge carriers will thus become apparent only when the cyclotron orbit is smaller than the characteristic puddle size, $l_c=\sqrt{\frac{\hbar}{eB}}=\textrm{ 25.64 nm}/\sqrt{B}\approx d/2$, where B is the magnetic field normal to the layer. This defines a characteristic field, $B_c=\frac{4\hbar}{ed^2}$, below which the LL spectrum is smeared out by disorder. Consistent with this picture we find no distinct LL features in the tunneling spectrum for fields below B$_c\approx$ 6 T (Fig. \ref{fig:fig3}(a)), corresponding to l$_c$(6 T)=10.5 nm $\approx$ d/2. Above 6 T the spectra develop peaks that evolve and become more pronounced with increasing field as shown in Fig. \ref{fig:fig3}(b). To compare the field and level index dependence of the measured spectra to that expected  for the quantized LL energies specific to the two dimensional Dirac electrons in graphene:
\begin{equation}
E=\pm v_F\sqrt{2e\hbar\left|N\right|\cdot B}, N=0,\pm1,\pm2,...
\label{eq:LLsinglelayer}
\end{equation}
we plot in Fig. \ref{fig:fig3}(c) the measured peak energies  as a function of the reduced parameter $sgn(N)\cdot\sqrt{\left|N\right|B}$. We find a linear dependence in agreement with the expected LL sequence for single layer graphene and from the linear fit we extract the Fermi velocity: v$_F$=(1.07 $\pm$ 0.02) $\times$ 10$^6$ m/s. We find that there is a 5-10$\%$ variation in the Fermi velocity across the  sample. This value is consistent with other measurements of the Fermi velocity in graphene on SiO$_2$ \cite{zhang2009origin,jiang2007infrared,henriksen2010interaction,martin2009nature,PhysRevLett.105.136801}. Compared to the results obtained for graphene on the surface of graphite \cite{PhysRevLett.102.176804} and for epitaxial graphene on SiC \cite{miller2009observing}, the levels are broader and their width varies with position suggesting a shorter carrier lifetime due to extrinsic scattering mechanisms such as trapped charges, ripples, etc.  The typical line-width for LLs with N$\neq$0, 20-30 meV, corresponds to carrier lifetimes, $\tau\approx \frac{\hbar}{\Delta E}, \tau\approx$ 22 fs - 32 fs consistent with values found by other methods in non-suspended graphene samples \cite{martin2007observation,jiang2007infrared,henriksen2010interaction,PhysRevLett.105.136801}.

\begin{figure}
\includegraphics[width=\columnwidth]{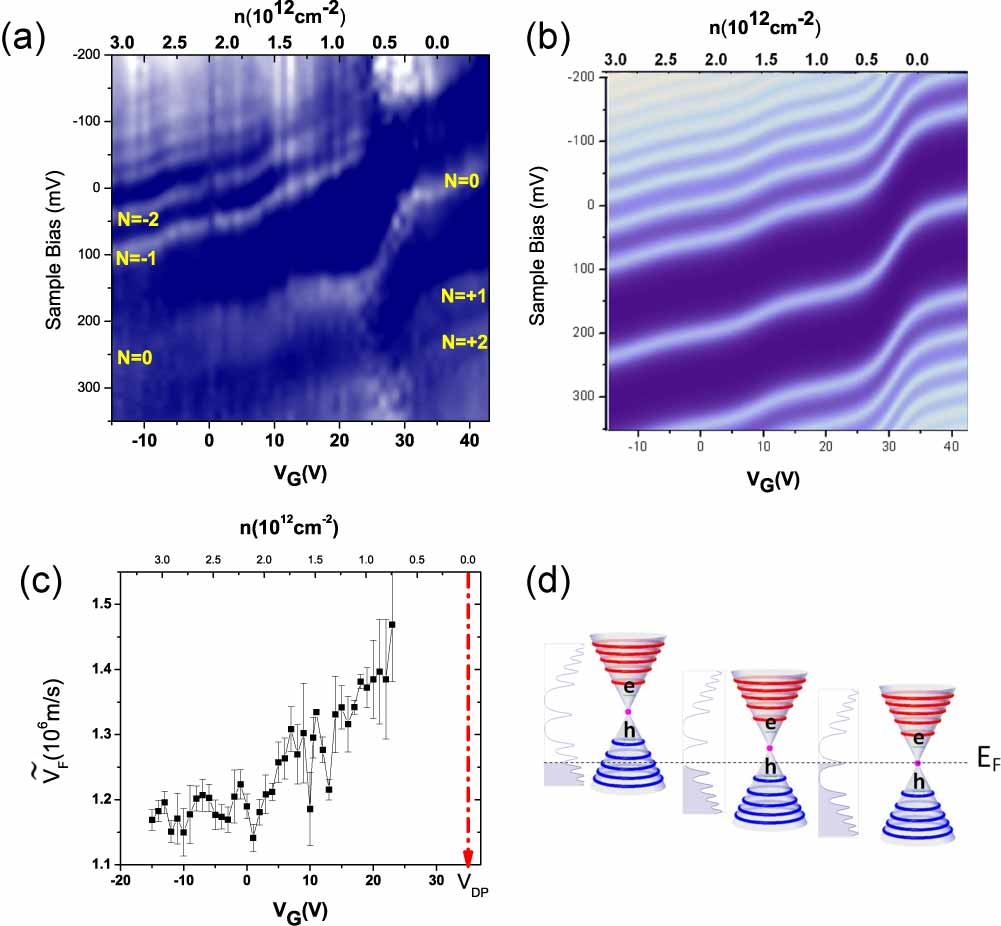}
\caption{\label{fig:fig4} (a) Measured STS as function of gate voltage V$_G$ (carrier density n). Each vertical line in the map is a spectrum at a particular gate. The bright stripes correspond to Landau Levels as indicated by N=0,$\pm$ 1,$\pm$ 2. (b) Simulation of the evolution of the LL spectrum with gate voltage, assuming Lorenzian line shapes of width $\approx$ 30 meV, for all levels and v$_F$=(1.16 $\pm$ 0.02)$\times$ 10$^6$ m/s. The latter was obtained from fitting the LLs at V$_G$=0 V , and C=100 aF$\mu$m$^{-2}$ for the capacitance between the graphene and the Si back gate. This data and that  in Fig. 3 were taken in  different parts of the sample, therefore the small variation in Fermi velocity. (c) Gate voltage dependence of $\tilde{V_F}$, the average slope of the dispersion. The Dirac point, at V$_G$= 35 V, is marked. (d) Illustration of the effect of gating on the Fermi level.}
\end{figure}

We note that the LL at the Dirac point (DP) where  N=0, is $\approx$ 200 meV above the Fermi level, indicating that the system is hole doped even though the gate is at ground potential. The corresponding carrier density, n=2 $\times$ 10$^{12}$ cm$^{-2}$ is found using: $\left|E_F-E_D\right|$=$\hbar v_F\sqrt{\pi n}$. This residual doping level, typical to graphene samples on SiO$_2$, can be attributed to charge impurities trapped between graphene and the SiO$_2$ or in the oxide. The energy of the DP relative to the Fermi level exhibits a small oscillation with field which, as discussed below, reflects the pinning of the Fermi level within each LL as the filling factor changes. This effect is much more pronounced when the filling factor is varied through gating as shown in Fig. \ref{fig:fig4}.

The absence of LLs in earlier STS measurements of graphene on SiO$_2$ despite evidence of their existence inferred by less direct probes \cite{jiang2007infrared,henriksen2010interaction,martin2009nature,PhysRevLett.105.136801} was puzzling \cite{geringer2010electrical}. To shed light on this question, we performed the same measurement on samples in which the SiO$_2$ dielectric was not chlorinated. The higher concentration of trapped charges in the unchlorinated substrates leads to a stronger random potential which is responsible for charge inhomogeneity and electron hole puddles in graphene. For these samples, even in those cases were peaks resembling LLs could be observed, their field and level index dependence did not follow the scaling expected for single layer graphene. Furthermore, similarly to earlier reports \cite{geringer2010electrical}, we encountered gating hysteresis where sweeping the gate back and forth did not give identical spectra. We conclude that in order to access the intrinsic properties of the charge carriers in graphene it is essential to minimize the effect of trapped charges by using an appropriately treated dielectric substrate.

The V shaped density of states in graphene gives rise to an ambipolar electric field effect which allows tuning the charge carrier density from electrons to holes by applying a gate voltage. Thus, for the samples studied here which are initially hole doped it is possible, by applying a positive gate, to bring the DP towards E$_F$, as illustrated in Fig. \ref{fig:fig4}(d). In the STS experiment by changing the sample tip bias we probe the occupied and unoccupied states above and below E$_F$. Therefore, when presenting the data it is convenient to keep the energy origin at E$_F$ so that the energy of the DP varies with respect to it. To study the effect of gating on the LLs we record the differential conductance spectra, dI/dV(E), at a fixed value of the magnetic field for a sequence of gate voltages. Our results for B=12 T are summarized as a map of the spectra versus gate voltage in Fig. \ref{fig:fig4}(a). Each vertical line in the map corresponds to the spectrum taken at a particular gate voltage. At V$_G$=-15 V, the DP (N=0) is at $\approx$ E=240 meV, followed by a sequence of clearly resolved LL N = -1, -2, -3 etc. We vary the gate voltage between: -15 $V < V
_{Gate} < $+43 V, corresponding to carrier densities: 3$\times 10^{12}$ cm$^{-2}$ $> n >$ -0.5 $\times 10^{12}$ cm$^{-2}$.

Several features in this data stand out. Varying the carrier density through gating is accompanied by pinning of E$_F$ to each LL as it is filled and followed by a jump of E$_F$ to the next LL when a level is full. Qualitatively, one can understand the step-like features as follows:  The LL spectrum consists of  peaks where  the DOS is large separated by minima  with low DOS. It takes a large change in the charge carrier density to fill the higher DOS regions,  resulting in  plateaus   where the Fermi level is ``pinned'' to a  particular Landau level. In contrast,  filling the low DOS region  in between the LLs does not require a large  change in carrier density –therefore the jumps (changes in slope) in between plateaus.  The broader the Landau levels are, the larger the slope in the plateau region and the  less abrupt the jumps in between. A similar effect was previously observed in very high mobility GaAs samples using time domain capacitance spectroscopy (TDCS)\cite{dial2007high}. Unlike the case of the 2DEG, due to the fact that in graphene the LL are not equally spaced, the largest jump from N=-1 to N=0 is followed by successively smaller jumps for higher index levels. Moreover, the broader the levels, the less abrupt the jumps, meaning that in disordered samples, this pinning effect is smeared out. 

 To analyze the data we model the Landau levels as Lorenzians with equal width $\approx$ 30 meV  from which we numerically calculate the chemical potential as a function of carrier density.The result of this calculation shown in Fig. \ref{fig:fig4}(b) is consistent with the experimental data. We note that as N=0 is brought closer to the Fermi level it becomes sharper and better defined as expected from theoretical considerations of minimal scattering at the Fermi energy \cite{gonzalez1999marginal}.
 
For a linear dispersion (Dirac cone) one can obtain the Fermi velocity by plotting the LL energies against $sgn(N)\cdot\sqrt{\left|N\right|B}$ and fitting to Equation \ref{eq:LLsinglelayer}. However when the dispersion is not strictly linear, as expected in the presence of electron-electron interactions, this procedure gives the average slope of the dispersion,$\tilde{V_F}$. Using this procedure for a series of gate voltages we found a substantial increase in $\tilde{V_F}$upon approaching the Dirac point, Fig. \ref{fig:fig4}(c),  consistent with the expected interaction-induced renormalization of the Dirac cone \cite{polini2007graphene,gonzalez1999marginal}. 
 
By using STM/STS in magnetic field we have shown here that for graphene supported on chlorinated SiO$_2$ substrates, the substrate disorder is weak enough to allow observation of quantized LLs already at moderate fields. The ability to control the doping level in this system made it 
possible to observe the interaction-induced  renormalization of the Dirac
cone at low carrier densities. These results show that intrinsic phenomena reflecting correlation effects between the massless Dirac fermions are observable and become apparent even in non-suspended samples for sufficiently clean substrates.

In a recent post Elias et al. \cite{Elias} report up to a three-fold interaction-induced  increase in  the Fermi velocity at low carrier densities obtained from the temperature dependence of the SdH in suspended graphene devices.

Work supported by DOE under DE-FG02-99ER45742 and partially supported by NSF-DMR-0906711, IAMDN and by Lucent. 

We thank M. Polini, A. MacDonald, A.K. Geim and F. Guinea  for useful discussions.

\end{document}